# Hard magnetic properties in nanoflake van der Waals Fe$_3$GeTe$_2$

Cheng Tan[1,#], Jinhwan Lee[2,#], Soon-Gil Jung[3], Tuson Park[3], Sultan Albarakati[1], James Partridge[1], Matthew R. Field[1], Dougal G. McCulloch[1], Lan Wang[1,*] & Changgu Lee[4,*]

[1]School of Science, RMIT University, Melbourne, VIC 3001, Australia.
[2]School of Mechanical Engineering, Sungkyunkwan University, Suwon, Republic of Korea.
[3]Center for Quantum Materials and Superconductivity (CQMS) and Department of Physics, Sungkyunkwan University, Suwon, Republic of Korea.
[4]School of Mechanical Engineering and SKKU Advanced Institute of Nanotechnology (SAINT), Sungkyunkwan University, Suwon, Republic of Korea.
[#] Equally contribute to the paper
[*] Corresponding authors. Correspondence and requests for materials should be addressed to L.W. (email: lan.wang@rmit.edu.au) and C. L. (email: peterlee@skku.edu.kr)

**ABSTRACT:** Two dimensional (2D) van der Waals (vdW) materials have demonstrated fascinating optical, electrical and thickness-dependent characteristics. These have been explored by numerous authors but reports on magnetic properties and spintronic applications of 2D vdW materials are scarce by comparison. By performing anomalous Hall effect transport measurements, we have characterised the thickness dependent magnetic properties of single crystalline vdW Fe$_3$GeTe$_2$. The nanoflakes of this vdW metallic material exhibit a single hard magnetic phase with a near square-shaped magnetic loop, large coercivity (up to 550 mT at 2 K), a Curie temperature near 200 K and strong perpendicular magnetic anisotropy. Using criticality analysis, we confirmed the existence of magnetic coupling between vdW atomic layers and obtained an estimated coupling length of ~ 5 vdW layers in Fe$_3$GeTe$_2$. Furthermore, the hard magnetic behaviour of Fe$_3$GeTe$_2$ can be well described by a proposed model. The magnetic properties of Fe$_3$GeTe$_2$ highlight its potential for integration into vdW magnetic heterostructures, paving the way for spintronic research and applications based on these devices.

Two dimensional (2D) van der Waals (vdW) materials have received considerable attention since the successful isolation of graphene[1, 2]. Studies on these materials have revealed novel optical[3, 4] and electronic[5, 6] properties. Moreover, employing heterostructures based on these 2D vdW materials has revealed further interesting properties and suggested applications [7, 8, 9]. vdW magnets were known more than 50 years ago[10, 11, 12], but interest has been renewed with the emergence of 2D materials. In the last few years, Raman spectroscopy[13, 14, 15, 16, 17] and electron transport measurements[18, 19] have been performed on 2D magnets. Importantly, 2D ferromagnetism has been discovered very recently in two insulating vdW materials, $Cr_2Ge_2Te_6$[20] and $CrI_3$[21] and novel devices based on vdW ferromagnetic heterostructures have been demonstrated[18, 22, 23]. The opportunity exists to design and fabricate many devices based on vdW magnets. For example, vdW magnetic insulator can magnetize 2D topological insulators by the magnetic proximity effect and thereby generate the quantum anomalous Hall effect in these materials[24, 25, 26, 27]. vdW ferromagnetic metals can be employed in spin orbit torque devices when stacked with vdW metals with strong spin orbit interactions[28, 29, 30, 31, 32, 33]. However, in order to exploit ferromagnetic vdW materials as building blocks for vdW heterostructure based spintronics, a ferromagnetic vdW metal with a hard magnetic phase and a large magnetic remanence to saturated magnetization ($M_R/M_S$) ratio is essential. This kind of vdW ferromagnetic metal is scarce.

Among all the predicted and experimentally observed vdW ferromagnetic materials[19, 20, 21, 23, 34, 35, 36, 37, 38, 39, 40, 41, 42, 43], a very promising ferromagnetic metal is $Fe_3GeTe_2$ (FGT), which exhibits a Curie temperature ($T_C$) near 220 K in its bulk state[42]. Previous experimental work has shown that bulk single crystalline FGT has a ferromagnetic state with a very small $M_R/M_S$ ratio and coercivity at all temperatures[42, 43, 44, 45], suggesting limited potential as a building block for vdW magnetic heterostructures. However, the $M_R/M_S$ ratio and coercivity of a magnetic material strongly depend on its domain structure[46, 47, 48], which is thickness dependent. Furthermore, recent research[49] shows that MBE-grown wafer-scale FGT thin films have improved magnetic properties. These findings motivate us to investigate the magnetic properties of exfoliated FGT nanoflakes of various thicknesses using anomalous Hall effect measurements.

Here, we report anomalous Hall effect measurements on single crystalline FGT nanoflakes and show that their magnetic properties are highly dependent on thickness. Importantly, by reducing the thickness to less than 200 nm, a hard magnetic phase with large coercivity and near square-shaped hysteresis loop occurs. These characteristics are accompanied by strong perpendicular magnetic anisotropy, making vdW FGT a ferromagnetic metal suitable for vdW heterostructure-based spintronics. By employing criticality analysis, the existence of magnetic coupling with a coupling length of ~ 5 vdW layers between vdW atomic layers is estimated in FGT. Finally, we propose a model to describe the hard magnetic behaviour of FGT thin flakes. This model is suitable for other vdW ferromagnetic thin films or nanoflakes with strong perpendicular anisotropy and square-shaped magnetic loops.

In a ferromagnetic material, the relationship between the Hall resistance and the applied magnetic field is given by Eq. 1.

$$R_{xy} = R_0 B_Z + R_S M_Z \qquad (1).$$

Here, $R_{xy}$ is the Hall resistance, which is composed of a normal Hall resistance (the first term in Eq. 1) and an anomalous Hall resistance (the second term in Eq. 1). $B_Z$ and $M_Z$ are the applied magnetic field and the sample magnetic moment perpendicular to the sample surface, respectively. The anomalous Hall resistance is proportional to $M_Z$. As FGT is a metallic ferromagnetic material, the normal Hall resistance is negligible compared with the anomalous Hall resistance in the magnetic field range of interest. Hence, the shape of the $R_{xy}$ vs B loop is actually the same as that of the $M_Z$ vs B loop. The coercivity and $M_R/M_S$ ratio can be obtained from the $R_{xy}(B)$ curve.

We measured the longitudinal resistance $R_{xx}$ and transversal resistance $R_{xy}$ of 11 FGT nanoflake devices with thickness from 5.8 nm to 329 nm and a bulk FGT crystal. The $R_{xy}(B)$ of selected FGT nanoflake devices at 2 K are shown in Fig. 1 (b-f) with the $R_{xy}(B)$ of the bulk FGT crystal (Fig. 1a). The applied magnetic field was perpendicular to the surfaces of the samples. The coercivity and $M_R/M_S$ ratio of the bulk crystal sample at 2 K are only 21.6 mT and 0.07, respectively. These characteristics agree well with those measured using a magnetometer[42]. However, the exfoliated nanoflakes displayed different magnetic properties. The nanoflakes with thicknesses of 329 nm and 191 nm displayed magnetic loops resembling two magnetic phases with different coercivities. As shown in Fig. 1b and 1c, when the magnetic field sweeps from the positive saturation field to the negative saturation field, the $R_{xy}$ values decrease sharply at a lower negative magnetic field and then decrease again at a higher negative field, which is similar to the behaviour of the coexistence of two phases. When the thicknesses of the FGT nanoflakes decrease further, the "two phase" behaviour disappears. As shown in Fig. 1(d-f), the $R_{xy}(B)$ loops of the three samples (with thicknesses of 82 nm, 49 nm and 10.4 nm, respectively) display a near square shape, indicative of a single hard magnetic phase. The coercivities of these three samples are much larger than those of the samples with 329 nm and 191 nm thickness and exceed 400 mT at 2 K. As FGT gradually evolves from a soft phase (bulk) to a single hard phase (82 nm, 49 nm and 10.4 nm), we speculate that the "two phase" behaviour in the nanoflakes with thicknesses of 329 nm and 191 nm is due to the gradual evolution of the domain structure. The $M_R/M_S$ ratios of the 191 nm, 82 nm, 49 nm and 10.4 nm thickness nanoflakes are near 1 at 2 K, demonstrating that all their magnetic moments remain aligned perpendicular to the sample surfaces at the remanence point. The magnetic moments flip abruptly to the opposite direction at the coercive field. In the magnetic field regime away from the coercivity, the four nanoflakes behave like a single magnetic domain with a strong perpendicular anisotropy. The magnetic domains only appear and flip to the opposite direction near the coercivity. As bulk single crystalline FGT also shows a strong perpendicular anisotropy, the anisotropy should be induced by the crystalline field. With increasing temperature, FGT gradually evolves from ferromagnetic to paramagnetic state. The Curie temperature ($T_C$) of each FGT nanoflake device can be determined from the temperature point where the remanence value goes to zero[20], as shown in the supplementary materials. The dependence of $T_C$ on thickness (from 0.3 mm to 49 nm) is shown in Fig. 2, from which we conclude that the $T_C$ of the FGT nanoflakes is almost independent of thickness in this range. As shown in the inset of Fig. 2, the $T_C$ decreases sharply as the thickness decreases from 25 nm to 5.8 nm. This behaviour is similar to that in $Cr_2Ge_2Te_6$[20], but is different from the behaviour in $CrI_3$[21]. It should be emphasized that nine of the eleven devices were fabricated in ambient condition with an air exposure of ~7 mins. The other two ultra-clean devices were made in a glove box ($O_2 < 0.1$ ppm, $H_2O < 0.1$ ppm) with h-BN and PMMA covering. The two batches of devices show the same magnetic characteristics (details in supplementary materials). The theory of critical behaviour[50] reveals that the finite thickness of flakes limits the divergence of the spin-spin correlation length at the $T_C$. As FGT is an itinerant metallic system, its spin-spin coupling should extend for many atomic layers, even along the out of plane directions. The spin-spin coupling range along the out of plane direction can be fitted as shown in the inset of Fig. 2 using the formula[51]

$$1 - T_C(n)/T_C(\infty) = [(N_0 + 1)/2n]^\lambda \qquad (2),$$

where $T_C$ is the Curie temperature, n is the number of atomic layers of a flake, $N_0$ is the spin-spin coupling range, and λ is a universal critical exponent. A best fitting to the data requires λ = 1.66 ± 0.18 and $N_0$ = 4.96 ± 0.72 monolayers. The fitted λ = 1.66 is near the value of a standard 3D Heisenberg ferromagnetism[52]. The correspondence achieved using a single fitting curve also indicates that FGT nanoflakes with a thickness of more than 5 vdW layers are still 3D ferromagnets. If FGT nanoflakes evolve from 3D ferromagnetism to 2D ferromagnetism from 25 nm to 5.8 nm, we should be able to obtain two fitting curves with different critical exponents. However, the data does not show this behaviour, which further confirms 3D magnetism when FGT thickness > 5.8 nm. Scaling behaviour[40, 41] near the $T_C$s of samples with thicknesses from monolayer to > 10 nm should reveal the evolution of the magnetism from 3D to 2D with decreasing thickness in FGT. As the focus of this paper is revealing the hard magnetic properties of FGT nanoflakes and their suitability for future spintronic applications, we propose this scaling analysis as future work.

More detailed measurements were performed on the 10.4 nm thickness nanoflake. In Fig. 3a and Fig. 3b, the $R_{xy}$ (B) loops from this sample, measured with perpendicular applied magnetic field, are plotted at various temperatures. There is a clear evolution with increasing temperature. At 2 K, the $R_{xy}$(B) loop is nearly square-shaped with a large coercivity of 552.1 mT and $M_R/M_S$ ~ 1, revealing alignment of spins due to strong perpendicular anisotropy. The $R_{xy}$ (B) loops remain approximately square up to 155 K. Figure 3d displays the temperature dependence of coercivity in this temperature regime. When the temperature exceeds $T_C$ (~191 K), the nanoflake becomes paramagnetic.

We also measured the temperature dependence of the $R_{xy}$ at the remanence point of the sample (the measurement and data process details of $R_{xy}(T)$ are shown in the method). The $R_{xy}(T)$ of the 10.4 nm nanoflake is shown in Fig. 3c. $R_{xy}(0)$ is an extrapolation to T = 0 K from region II based on the spin wave model as discussed later. Results from other samples are shown in the supplementary materials. Below 155 K, the FGT nanoflake exhibits a ferromagnetic phase with near square-shaped magnetic loop. The abrupt decrease of the magnetic moment around 155 K in Fig. 3c indicates a first order magnetic phase transition not yet known, but likely related to the competition between the perpendicular anisotropic energy and the thermal agitation energy. As the bulk FGT single crystal also shows a phase transition with gradually changed magnetic moment around 155 K[45], we speculate that the sharper phase transition in the FGT nanoflake is due to its decreased thickness, which induces single domain behaviour at the remanence. In the temperature regime from 155 K to $T_C$, the FGT nanoflake displays a ferromagnetic phase with very small coercivity and remanence. The $R_{xy}(T)$ reveals another phase transition near 10 K, where the remanence increases sharply with decreasing temperature, indicative of the formation of new spins and magnetic moments. Further understanding of the phases present in this temperature regime would require neutron scattering measurements, which are beyond the scope of this article. Fig. 3c also shows the temperature dependence of $R_{xy}(T)$ fitted from 2 K to 150 K using the mean field theory (the Brillouin function) and the spin wave theory. The $R_{xy}(T)$ behaviour of the 10.4 nm thickness sample cannot be fitted using mean field theory, but agreement with the spin wave theory for a three-dimensional ferromagnet is good. This provides further evidence that a FGT nanoflake remains a 3D magnetic system when its thickness exceeds 5 layers. Our experimental results contain information required to construct a correct model for FGT. First, as shown in $R_{xy}(B)$ measurements (Fig. 1, Fig. 3 and Fig.4), FGT has a very strong perpendicular anisotropy due to the crystalline field. Second, the thickness dependence of $T_C$ as shown in Fig. 2 indicates the existence of magnetic coupling between atomic layers in FGT with a coupling range of about 5 vdW layers. Therefore, a correct Hamiltonian describing FGT should include a perpendicular anisotropic energy, an in-plane spin-spin interaction energy, an out of plane spin-spin interaction energy and a Zeeman energy induced by the applied magnetic field.

The evolution of $R_{xy}$ hysteresis loops with the angle $\theta$ between the applied magnetic field and the direction perpendicular to the sample surface (i.e. the direction of magnetic anisotropy) for the 10.4 nm flake at 2 K is shown in Fig. 4a and 4b. As mentioned prior, the spins in the FGT nanoflakes align to one direction due to the strong perpendicular anisotropy when the temperature is below 155 K. Magnetic domains only appear near the coercive field. This simple magnetic structure makes it possible to construct a model to describe the behaviour of the coercivity. When a magnetic field is applied to a single domain ferromagnetic material with uniaxial anisotropy, the energy of the magnetic system is composed of the magnetic anisotropic energy and the Zeeman energy,

$$E(T) = K_A(T)V_S \sin^2(\phi-\theta) - M_S(T)BV_S \cos\phi \quad (3),$$

where $K_A$ is the magnetic anisotropic energy per volume, $V_S$ is the volume of the sample, $\phi$ is the angle between the magnetic field and the magnetic moment, $\theta$ is the angle between the magnetic field and the direction of magnetic anisotropy (Fig. 4f), $M_S(T)$ is the magnetic moment of a unit volume FGT at temperature T, and B is the applied magnetic field. With an applied magnetic field B and the angle $\theta$, we can calculate $\phi$ from Eq. 4, the well-known Stoner-Wohlfarth model[53]

$$\frac{\partial E(T)}{\partial \phi} = 2K_A(T)V_S \sin(\phi-\theta)\cos(\phi-\theta) + M_S(T)BV_S \sin\phi = 0 \quad (4),$$

For thin FGT nanoflakes with perpendicular anisotropy, the demagnetization effect[46] should be considered. Consequently, the applied magnetic field B and angle θ in Eq. 3 and 4 should be modified to the effective magnetic $B_{eff}$ and angle $\theta_{eff}$. Further detail is shown in supplementary materials. Using this model, we fitted the $R_{xy}(B)$ curve with θ = 85° to obtain the unit magnetic anisotropic energy $K_A$ at 2 K, 25 K, 50 K, 80 K, 100 K, and 120 K, as shown in Fig. 4e (additional details provided in supplementary materials). Fig. 4c shows the fitting curve for the 2 K data. It should be emphasized here that all the $R_{xy}$ (B) with various θ at different temperatures in the magnetic regime away from the coercivity can be well fitted by the Stoner-Wohlfarth model. The reason for using θ = 85° loops for the $K_A$ fitting is that a magnetic loop of small θ value is nearly a straight line without curvature in the magnetic regime away from the coercivity, which is not suitable for obtaining a reliable $K_A$.

As magnetic domains appear in the magnetic field regime near coercivity, to describe the angular dependence of coercivity, the flip of magnetic domains near coercivity should be included in model. By considering the thermal agitation energy and utilizing the fitted $K_A$ values, a modified Stoner-Wohlfarth model (details in supplementary materials) can be used to describe the angular dependence of coercivity. As shown in Fig. 4g, if the system can be thermally excited from a meta-stable state (state 1) to an unstable state (state 2), the magnetic moment can then be flipped to a final stable state (state 3). The energy difference between state 1 and state 2 is ΔE. With increasing magnetic field B (more negative B), the energy difference ΔE between the stable state 1 and the unstable state 2 decreases. In the modified Stoner-Wohlfarth model, we make two assumptions:

1. At a certain B field, the thermal agitation energy is large enough to overcome the ΔE in a standard experimental time (100 seconds) causing the magnetic moment to flip. As the FGT shows a nearly square-shaped $R_{xy}$ loop (magnetization loop), we can assume that this B field is the coercive field, which is a reasonable approximation due to the sharp transition of the magnetic moments.

2. When the first domain flips under an applied magnetic field, other un-flipped magnetic moments will generate an effective field on the magnetic moment in the first flipped domain. The processes of domain flip, expansion, and interaction are complex. Micro-magnetic simulation is required to provide a detailed description, which is beyond the scope of this paper. Here, a parameter $a$(T) is used to describe the mean field interaction between the flipped and un-flipped magnetic moments.

Based on the proposal of Neel and Brown[54,55], we use $\Delta E = 25k_BT$ as the barrier height where the magnetic moment starts to flip (details in supplementary materials). We thus obtain

$$[K_A(T)V(T)\sin^2(\phi_2 - \theta) - [1 - a(T)]V(T)M_S(T)B\cos\phi_2] - \\ [K_A(T)V(T)\sin^2(\phi_1 - \theta) - [1 - a(T)]V(T)M_S(T)B\cos\phi_1] = 25k_BT \quad (5),$$

where V(T) is the volume of the first flipped domain at T, $a$(T) is the parameter describing the effective field due to the coupling between the first flipped domain and the un-flipped magnetic moments at temperature T, which affects the Zeeman energy. The value of $a$(T) lies between 0 and 1. $\phi_1$ and $\phi_2$ are the angles between the applied magnetic field and the magnetic moment for state 1 and state 2, respectively. These angles are calculated using Eq.4. Due to the demagnetization effect[46], B and θ here should also be modified to $B_{eff}$ and $\theta_{eff}$ (Supplementary materials).

Using V(T) and $a$(T) as fitting parameters in Eq.5 in conjunction with the modified Stoner-Wohlfarth model provides excellent agreement with the experimental angular dependence of coercivity at various temperatures (Fig. 4d), which further confirms the single domain behaviour induced by the strong perpendicular anisotropic energy in FGT in the field regime away from coercivity. The volume of the first flipped magnetic domain near the coercivity

V(T) is important for understanding the magnetic dynamics of a ferromagnetic materials. This volume V and the perpendicular anisotropic energy $K_A$ at different temperatures are shown in Fig. 4e. The modified Stoner-Wohlfarth model proposed here is suitable for describing the magnetic behaviour of 2D vdW ferromagnetic materials with strong perpendicular anisotropy and near square-shaped loop.

To conclude, FGT nanoflakes are vdW 2D metallic ferromagnets with large coercivity, $M_R/M_S$ ratio of 1, relatively high $T_C$ and strong perpendicular anisotropy. Exploitation of this new material in various vdW magnetic heterostructures with properties such as giant magnetoresistance and tunnelling magnetoresistance, as well as vdW spin-orbit torque heterostructures is expected to yield exciting results. This discovery paves the way for a new research field, namely, vdW heterostructure-based spintronics.

## Method

### Single Crystal Growth

Single crystal $Fe_3GeTe_2$ was grown by the chemical vapor transport (CVT) method. High-purity Fe, Ge and Te were blended in powder form with molar proportions of 3:1:5 (Fe:Ge:Te). Iodine (5 mg/cm$^2$) was added as a transport agent and the mixed constituents were sealed into an evacuated quartz glass ampoule. This ampoule was placed in a tubular furnace, which has a temperature gradient between 700~650 ºC. The furnace center temperature was ramped up to 700 ºC with a heating rate of 1 ºC per minute and was maintained at the set point for 96 hours. To improve crystallinity, the ampoule was slowly cooled down to 450 ºC for over 250 hours. Below 450 ºC, the furnace was cooled more rapidly to room temperature.

**Device Fabrication and Measurement**

First, the single crystalline $Fe_3GeTe_2$ was mechanically exfoliated and placed on a Si substrate with a 280 nm thickness $SiO_2$ layer. Then, Cr/Au (5 nm/100 nm) contacts were patterned by photolithography and e-beam lithography. During intervals between processing, the sample was covered by a PDMS film and stored in an evacuated glass tube (~$10^{-6}$ Torr). The sample was exposed to ambient for no more than 7 minutes throughout the fabrication procedure. The transport measurements were performed in a Quantum Design PPMS with 9 T magnetic field.

**Hall effect measurement and data processing**

Because of the non-symmetry in our nanoflake devices, the measured Hall resistance was mixed with the longitudinal magnetoresistance. We processed the data by using ($R_{xyA}$(+B) - $R_{xyB}$(-B)) / 2 to eliminate the contribution from the longitudinal magnetoresistance, where $R_{xyA}$ is the half loop sweeping from the positive field to the negative field, $R_{xyB}$ is the half loop sweeping from the negative field to the positive field, and B is the applied magnetic field. We also measured the $R_{xy}$ (T) at the remanence point for most of the samples. To measure the $R_{xy}$ (T) at the remanence point, the magnetic moment of samples was first saturated by a 1 T magnetic field and then the magnetic field was decreased to zero (the remanence point). Finally, the temperature dependence of the $R_{xy}$ at remanence was measured when the temperature was increased from 2 K to 300 K. In order to eliminate the non-symmetry effect of the device, we measured $R_{xy}$ (remanence) with both 1 T and -1 T saturation. The real $R_{xy}$ (T) at remanence without $R_{xx}$ mixing was calculated using ($R_{xyA}$(T) - $R_{xyB}$(T)) / 2. Here $R_{xyA}$ and $R_{xyB}$ are the remanence with 1 T and -1 T saturation, respectively.

# Associated content

Supplementary Materials, including the characterization of FGT crystals, detailed measurement data for other samples, definition of the Curie temperature, anomalous Hall effect and the modified Stoner-Wohlfarth model, mean field fitting and spin wave fitting, the effect of the surface amorphous oxide layer and the confirmation of ohmic contact.

# Acknowledgement

This research was supported by the Australian Research Council Centre of Excellence in Future Low-Energy Electronics Technologies (Project No. CE170100039), the Institute for Information & Communications Technology Promotion (IITP) grant (Project No. B0117-16-1003, Fundamental technologies of 2D materials and devices for the platform of new-functional smart devices), and the Basic Science Research Program (Project No. 2016R1A2B4012931), and the National Research Foundation (NRF) of Korea by a grant funded by the Korean Ministry of Science, ICT and Planning (Project No. 2012R1A3A2048816).

# Author Information

### Author notes

Cheng Tan and Jinhwan Lee contributed equally to this work.

### Affiliations

**School of Science, RMIT University, Melbourne, VIC 3000, Australia.**

Cheng Tan, Sultan Albarakati, James Partridge, Matthew R. Field, Dougal G. McCulloch & Lan Wang

**School of Mechanical Engineering, Sungkyunkwan University, Suwon, Republic of Korea**

Jinhwan Lee

**School of Mechanical Engineering and SKKU Advanced Institute of Nanotechnology (SAINT), Sungkyunkwan University, Suwon, Republic of Korea**

Changgu Lee

**Center for Quantum Materials and Superconductivity (CQMS) and Department of Physics, Sungkyunkwan University, Suwon, Republic of Korea.**

Soon-Gil Jung & Tuson Park

## Contributions

C.L. and L.W. conceived and designed the research. J.L. synthesized the material, J.L., S.J. and T.P. performed the material characterization. M.R.F. and D.G.M. performed the TEM scan for the cross-section of nanoflakes. C.T. J.P. and S.A. fabricated the Hall bar devices. C.T. and L.W. performed the electron transport measurements, data analysis and modeling. C.T., J.P., L.W., J.L. and C.L. wrote the paper with the help from all of the other co-authors.

## Competing interests

The authors declare no competing financial interests.

## Corresponding author

Correspondence to Changgu Lee: peterlee@skku.edu.kr

Lan Wang: lan.wang@rmit.edu.au

**Figure legends :**

**Figure 1.** $R_{xy}$(B) for FGT nanoflakes of various thicknesses at 2 K. Each red scale bars represents 10 μm. **a** Bulk device, L × W × T = 2 mm × 0.8 mm × 0.3 mm. $M_R/M_S$ = 0.0715. **b** A device with a thickness of 329 nm, L × W × T = 44.4 μm× 49.4 μm× 329 nm. $M_R/M_S$ = 0.0807. **c** A device with a thickness of 191 nm, L × W × T = 14.7 μm× 9.25 μm × 191 nm. $M_R/M_S$ = 0.9757. **d** A device with a thickness of 82 nm, L × W × T= 10.3 μm× 19.5 μm× 82 nm. $M_R/M_S$ = 0.9839. **e** A device with a thickness of 49 nm, L × W × T= 12.6 μm× 13.1 μm× 49 nm. $M_R/M_S$ = 0.9980. **f** A device with a thickness of 10.4 nm, L × W × T= 12.7 μm× 8.79 μm× 10.4 nm. $M_R/M_S$ = 0.9973.

**Figure 2.** Thickness dependence of the Curie temperature ($T_C$). Excepting the device with a thickness of 329 nm, the $T_C$s of the devices were determined by the temperature point when the remanence becomes zero. The $T_C$ of the device with a thickness of 329 nm was determined from its $\rho_{xx}$(T) curve. Inset shows the thickness dependence of the $T_C$s in nanoflakes with thicknesses from 5.8 nm to 25 nm with fitting curve. Blue dots are the $T_C$s for the two ultra-clean devices that were covered by h-BN and PMMA in a glove box ($O_2$ < 0.1 ppm and $H_2O$ < 0.1 ppm).

**Figure 3.** Anomalous Hall effect measurements performed on 10.4 nm thickness FGT device. **a** $R_{xy}$ (B) loops in the temperature regime from 2 K to 140 K, in which the FGT nanoflake shows ferromagnetic properties. **b** $R_{xy}$(B) curves in the temperature regime from 150 K to 185 K. The FGT nanoflake shows zero coercivity and remanence when T >150 K. **c** Normalized $R_{xy}$(T) curve and three fitting curves based on the mean field theory J = 1, J = ∞, and the spin wave theory, respectively. Three magnetic regimes exist from 2 K to $T_C$ = 191 K. Regime I (2~10K) is an unknown phase requiring further investigation. Regime II (10~155 K) is a hard ferromagnetic phase. Regime III (155~191 K) is a phase with zero coercivity and remanence. **d** The temperature dependence of coercivity from 2 K to 150 K.

**Figure 4.** Angular dependent Hall-effect measurements and relevant fittings using a modified Stoner–Wohlfarth model. **a, b** $R_{xy}$(B) loops at different angles between the applied magnetic field and the direction perpendicular to the surface of the nanoflake with a thickness of 10.4 nm. At 0°, the surface of the nanoflake is perpendicular to the magnetic field. **c** Normalized $R_{xy}$(B) curve at 2 K measured at 85° from 6 T to -6 T. The fitting curve is based on the Stoner-Wohlfarth model. **d** The effective angular dependence of effective coercivities at different temperatures. The solid curves are the fitting curves based on a modified Stoner-Wohlfarth model. From **b** we can tell that the remanences of the

$R_{xy}$ loops at angles > 85° show pronounced decreases and a divergent coercivity value was obtained at 90°, an angle beyond the range included by the modified Stoner-Wohlfarth model. Based on this experimental result, the theoretical fittings were only performed to 85°. **e** The temperature dependence of $K_A$ and V. $K_A$ was fitted by the Stoner–Wohlfarth model, while V was fitted based on the fitted $K_A$ and the modified Stoner-Wohlfarth model (more details in supplementary information). **f** Illustration of the variables used in the Stoner–Wohlfarth model. The dashed line is the easy axis of the magnetic anisotropy in FGT nanoflakes. **g** Schematic diagram of a magnetic system changing from a stable to an unstable state.

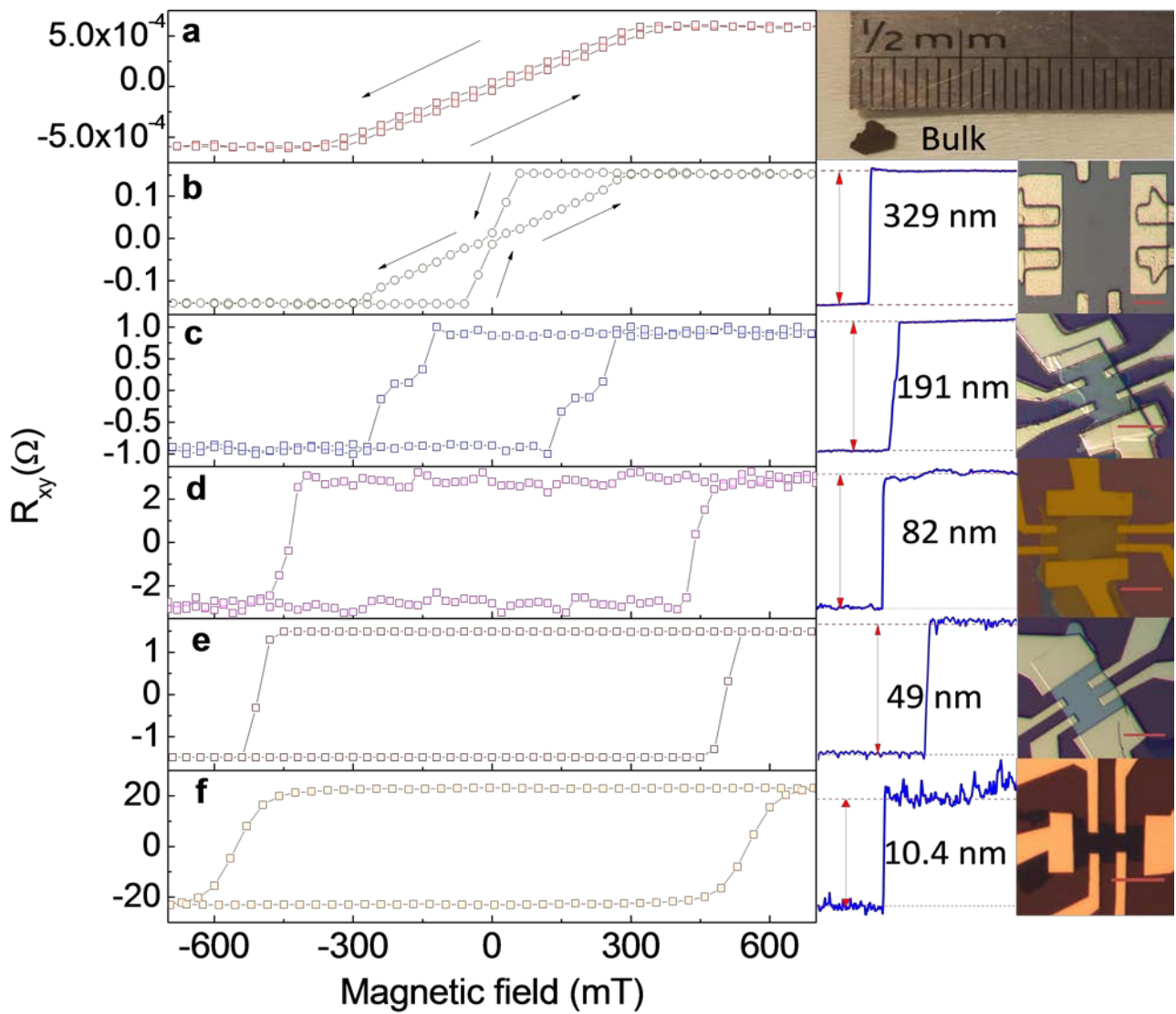

**Figure 1**

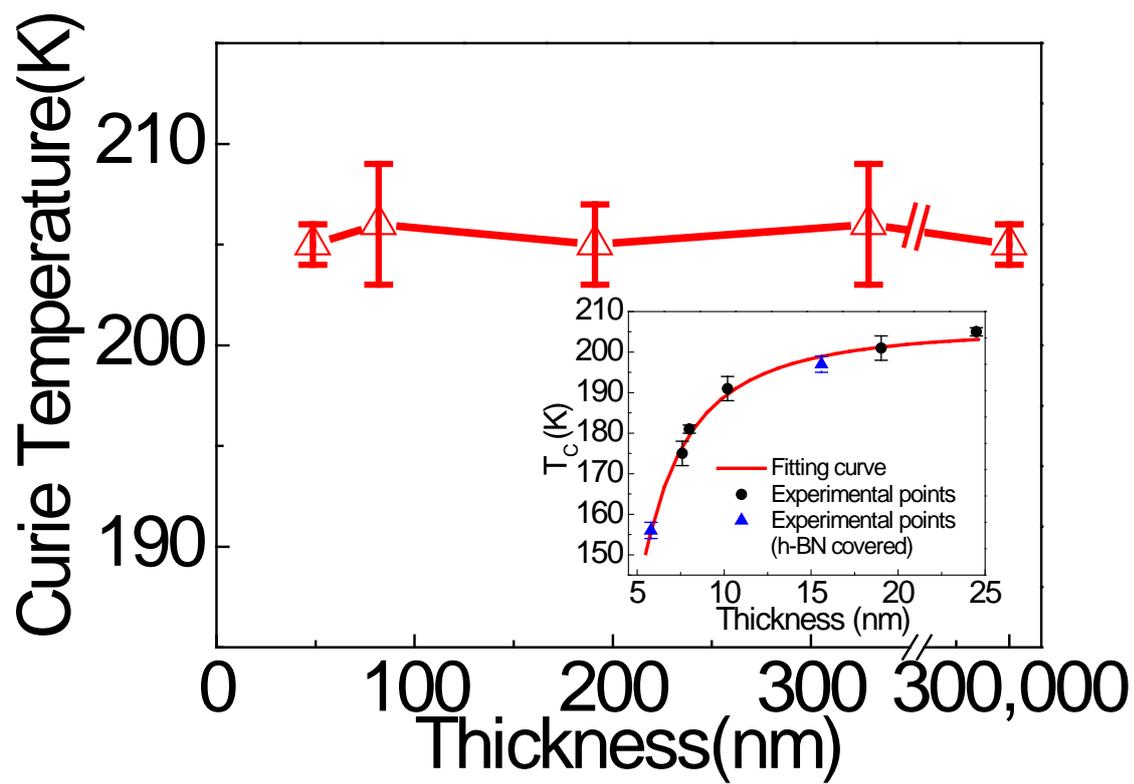

**Figure 2**

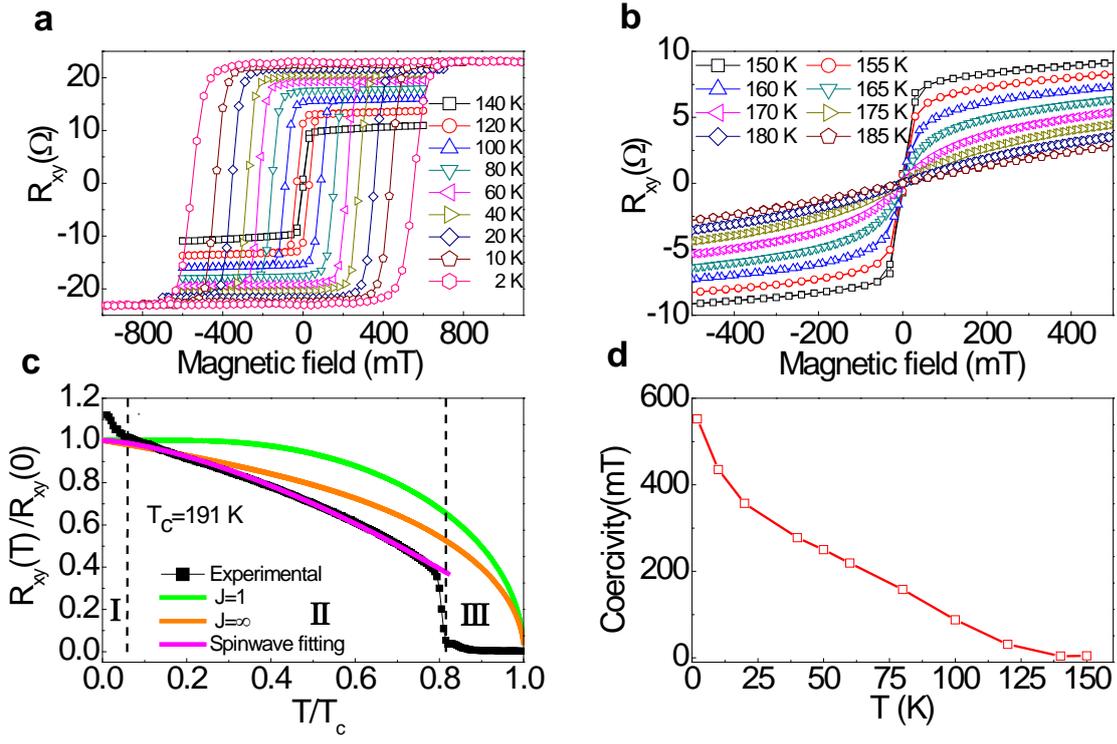

**Figure 3**

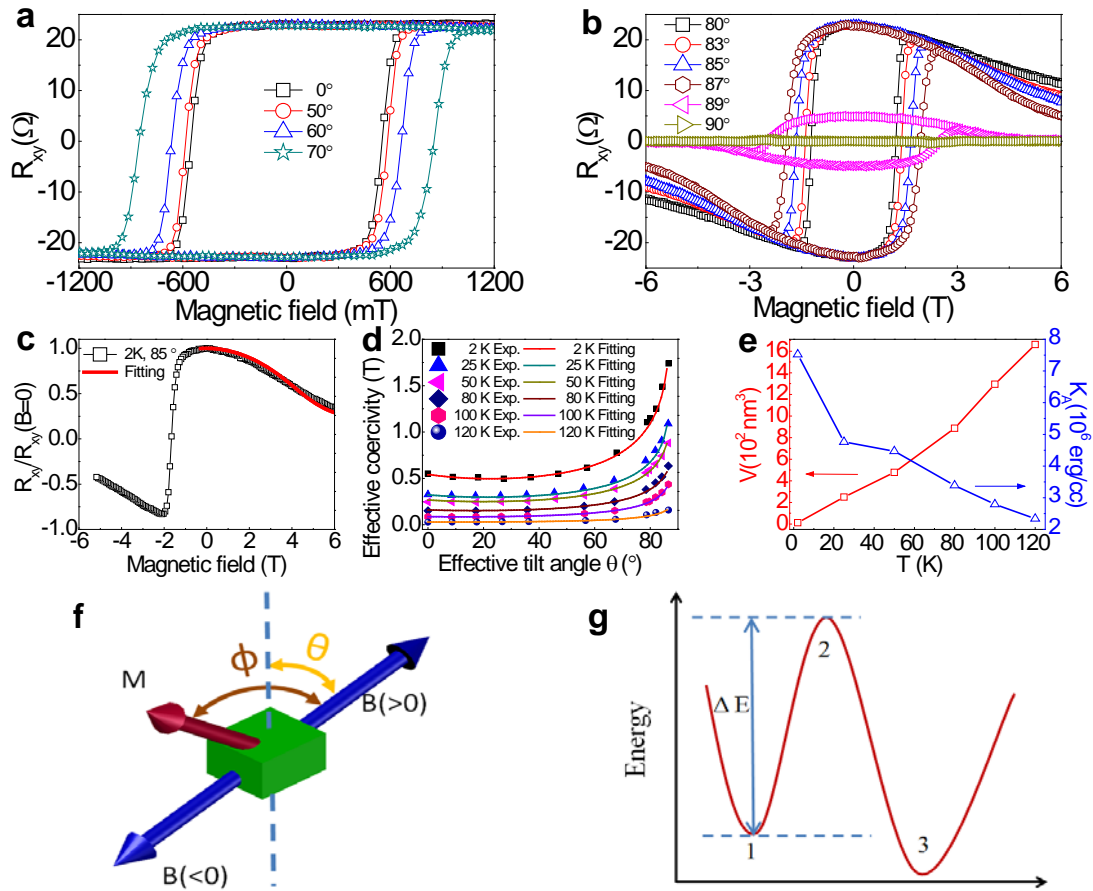

**Figure 4**

# Supplementary Materials

Section 1: The characterization of FGT crystals

Section 2: Detailed measurement data for other samples

Section 3: Definition of the Curie temperature

Section 4: Anomalous Hall effect and the modified Stoner-Wohlfarth model

Section 5: Mean field fitting and spin wave fitting

Section 6: The effect of the surface amorphous oxide layer

Section 7: The confirmation of ohmic contact

# Section 1: The material characterization of FGT crystal

**Section 1: The characterization of FGT crystals**

*Chemical composition*

To analyse the chemical composition of synthesized $Fe_3GeTe_2$ (FGT) flakes, energy-dispersive X-ray spectroscopy (EDS) was carried out. As shown in Fig. S1(a), we used bulk single crystalline FGT flakes after peeling off surface layers. EDS (Fig. S1(b)) revealed the FGT to be ~ 2.88:1:2.05 Fe:Ge:Te.

*Basic magnetic properties*

A magnetic property measurement system (MPMS), which includes a superconducting quantum interference device (SQUID), was employed to characterize the magnetic properties of FGT. The volume of sample was 0.00012329 $cm^3$. Figure S1(d) shows the temperature dependence of the inverse of magnetization. The data was collected in a magnetic field of 100 Oe, applied along the c-axis direction. We extracted the Curie temperature ($T_c$) from fitting according to the Curie-Weiss law in a magnetization graph, as shown in Fig. S1(d).

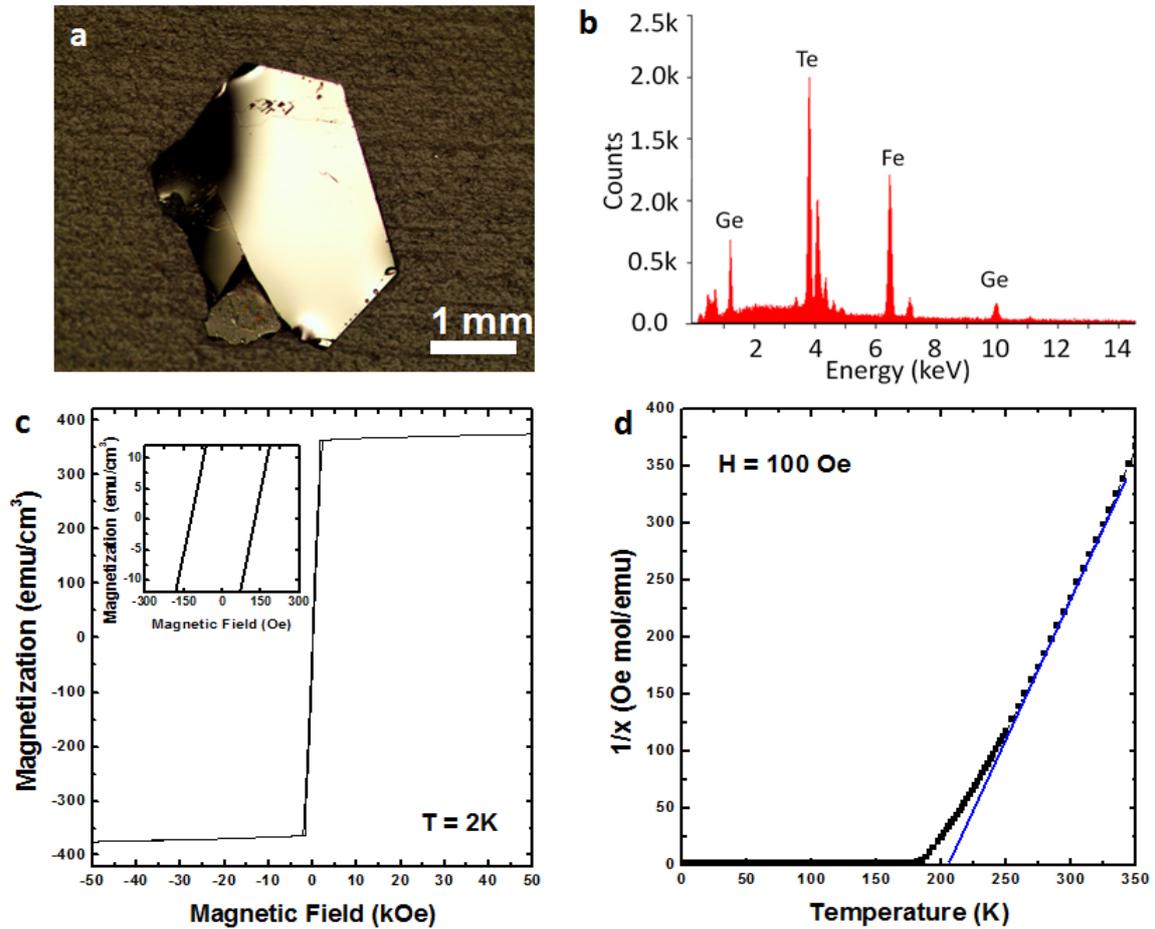

**Supplementary Fig. S1. Characterization of the FGT single crystal.** (a) Optical image of a synthesized FGT single crystal. (b) EDS spectrum of synthesized FGT single crystal. The stoichiometric composition of three elements Fe, Ge, and Te is 48.59: 16.83: 34.58 for the synthesized crystal. (c) Field dependence of magnetization at 2 K. The calculated Ms is 374.48 emu/cm$^3$. (d) Temperature dependence of inverse magnetization and fitted line for 100 Oe. The fitted line indicates the $T_C$ = 205 K.

# Section 2: Detailed measurement data for all the samples

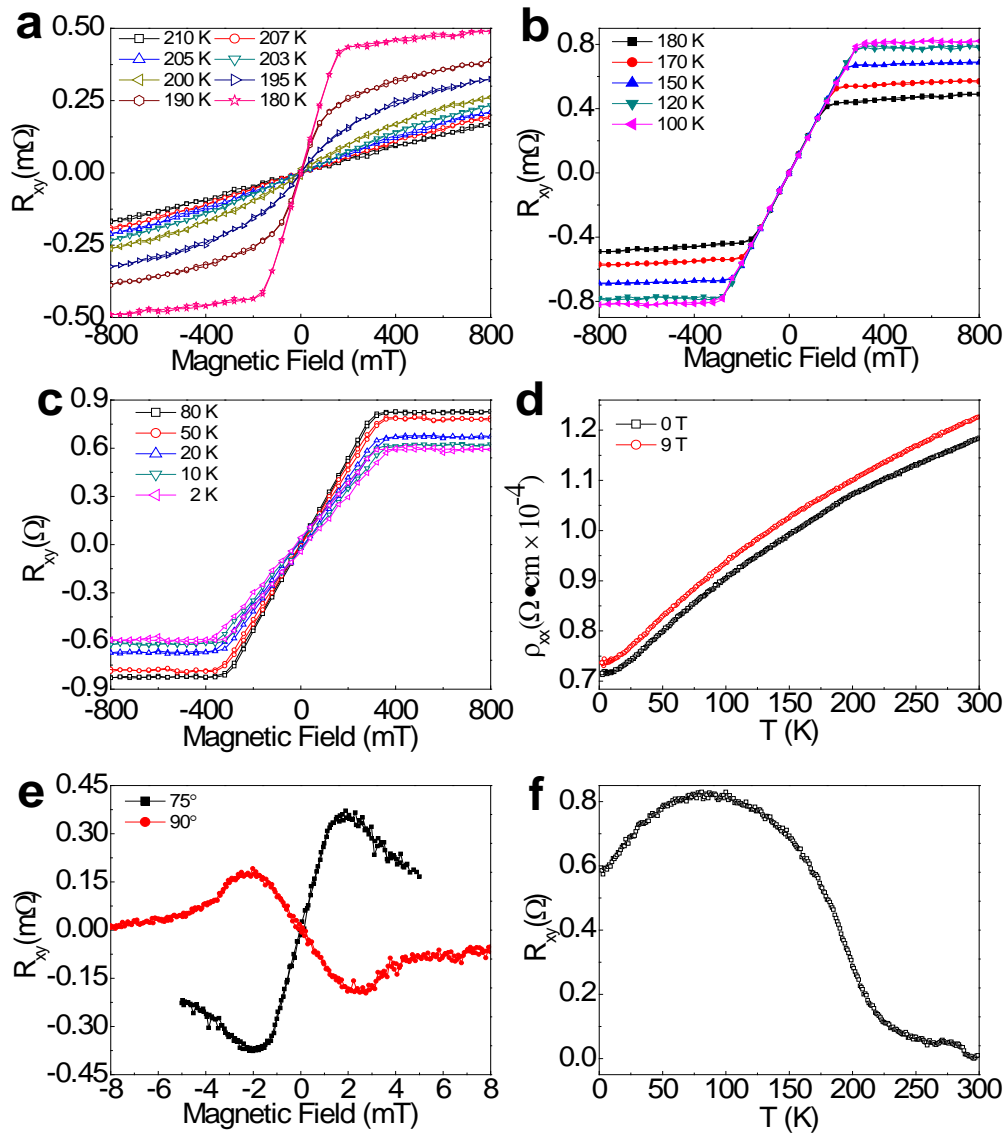

**Supplementary Fig. S2. $R_{xy}$ and $R_{xx}$ measurements for the bulk FGT single crystal.** The contacts for the bulk device were made of Indium. **(a)** $R_{xy}$ (B) curves show zero hysteresis in the temperature regime from 210 K to 180 K. The curve becomes linear at ~207 K. **(b)** $R_{xy}$ (B) loops show no hysteresis from 100 K to 180 K. **(c)** $R_{xy}$ (B) loops from 80 to 2 K. In this regime, the saturated magnetization decreases with decreasing temperature. At 2 K, the coercivity is ~21.6 mT. **(d)** Temperature dependence of $\rho_{xx}$ at 0 T and 9 T. **(e)** Angular dependent Hall effect at 50 K with θ = 75° and 90.3°, respectively. When a high in-plane magnetic field (θ = 90°) is applied, the electron spins with perpendicular anisotropy will be forced parallel to the sample plane and then point to random direction when the magnetic field is swept back to zero, therefore the $R_{xy}$ should be 0 in this condition. **(f)** $R_{xy}$ (T) curve under 1 T field.

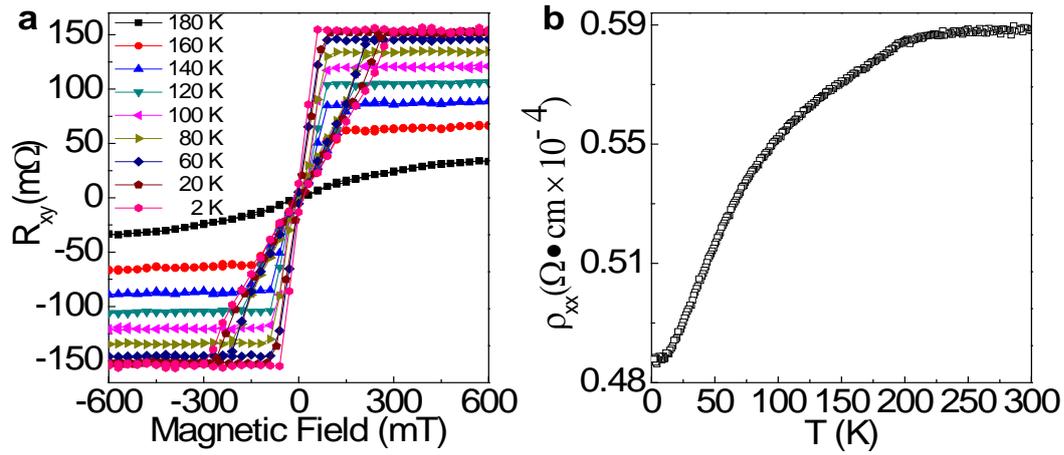

**Supplementary Fig. S3. $R_{xy}$(B) and $R_{xx}$(T) curves for the FGT nanoflake with a thickness of 329 nm.** (a) $R_{xy}$(B) curve at different temperatures. The hysteresis loops are not square. However, pronounced hysteresis loops with a non-zero coercive field are displayed. (b) $\rho_{xx}$(T) curve at zero magnetic field, an obvious magnetic transition due to spin-flip scattering is shown at about 205 K, from which the Curie temperature is determined.

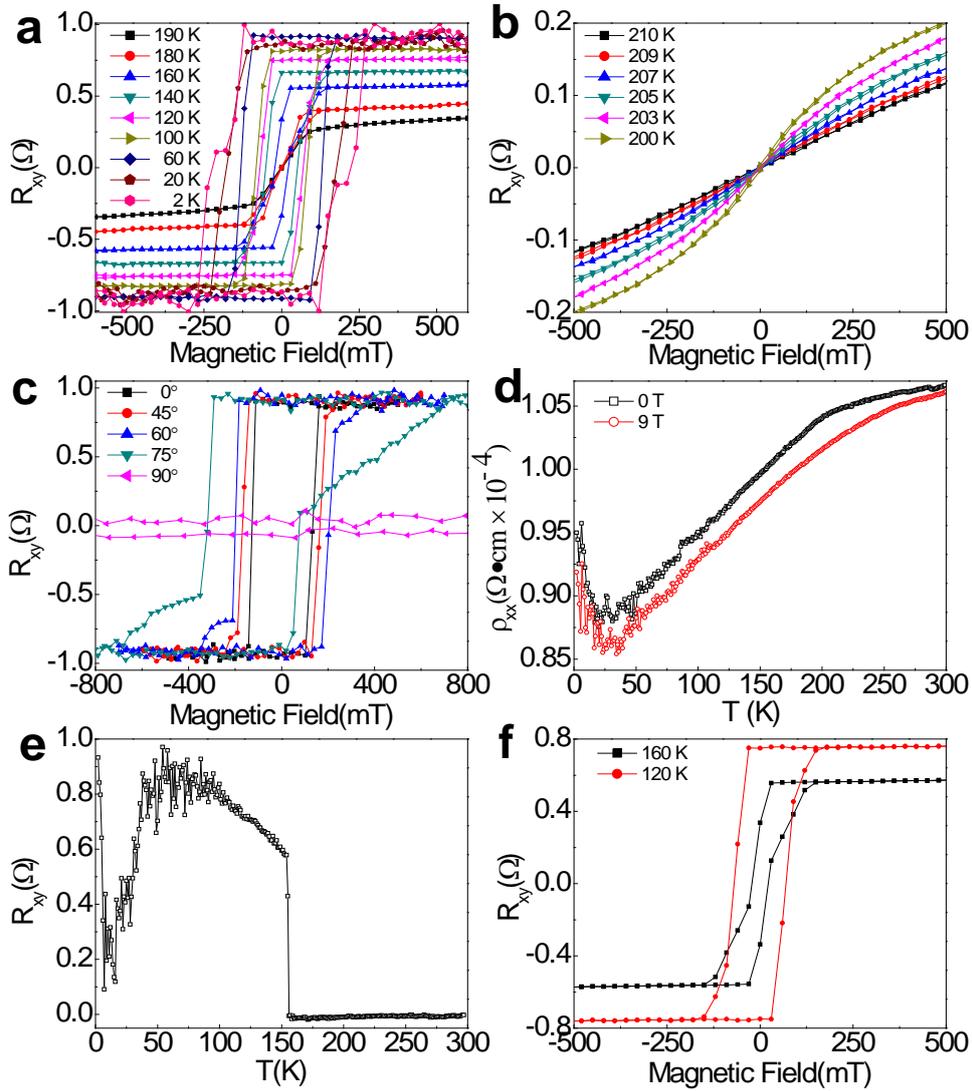

**Supplementary Fig. S4. $R_{xy}$ and $R_{xx}$ measurements for the nanoflake with a thickness of 191 nm.** (a) $R_{xy}$ (B) loops from 2 K to 190 K. The coercivity increases with decreasing temperature. (b) $R_{xy}$(B) curve from 200 K to 210 K. The curve becomes linear at ~209 K. All the curves shows no hysteresis. (c) Angular dependent Hall effect at 50 K, the $R_{xy}$ at 90° is nearly 0, which means the sample is nearly parallel to the magnetic field. At 75° an asymmetry loop is observed, which may indicate exchange coupling induced uni-directional magnetic anisotropy. (d) Temperature dependence of $\rho_{xx}$ at 0 T and 9 T. Although the data is noisy at lower temperature, the spin-flip scattering is still clear. (e) Temperature dependence of the remanence. The sudden decrease of remanence indicates that the thermo agitation energy is higher than the perpendicular anisotropic energy, which can also be clearly seen from the evolution of the $R_{xy}$ (B) loop from 160 K to 210 K. (f) $R_{xy}$(B) curve at 120 K and 160 K. Compared with the 120 K curve, the 160 K curve is indicative of two magnetic phases with increasing temperature.

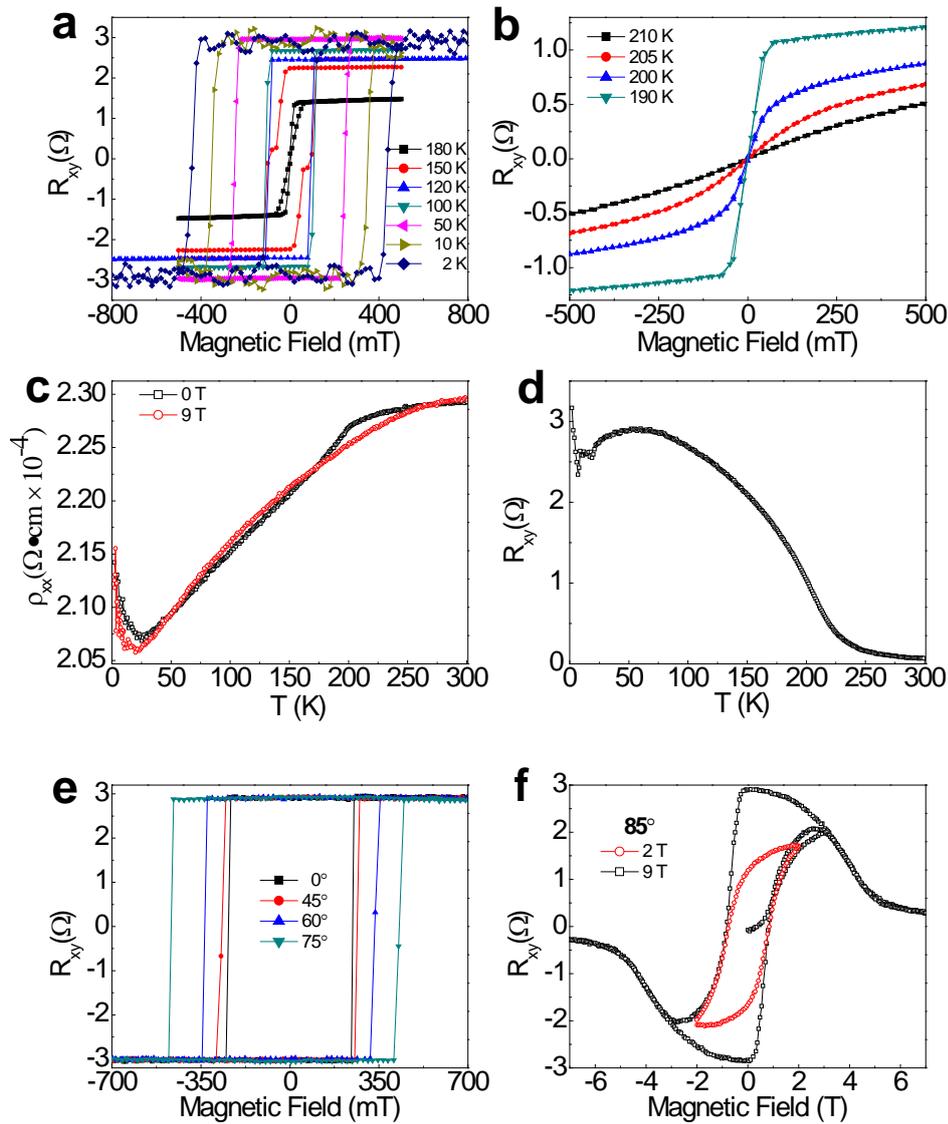

**Supplementary Fig. S5. $R_{xy}$ and $R_{xx}$ measurements for the nanoflake with a thickness of 82 nm. (a)** $R_{xy}$ (B) loops from 2 K to 180 K. The 150 K and 180 K loops show behaviour indicative of two magnetic phases. **(b)** $R_{xy}$ (B) curves from 190 K to 210 K. All curves do not show any hysteresis. **(c)** Temperature dependence of $\rho_{xx}$ at 0 T and 9 T. Spin-flip scattering happens at ~205 K. **(d)** $R_{xy}$ (T) curve under 1 T applied field. **(e)(f)** Angular dependent Hall effect at various angles θ at 50 K. **(f)** The angular dependent Hall effect at 85°. At first, the magnetic field was swept within 2 T, In this range, the curve did not show saturated behaviour. The $R_{xy}$ (B) shows saturated behaviour when the applied magnetic field exceeds 5 T, indicating a strong perpendicular anisotropy.

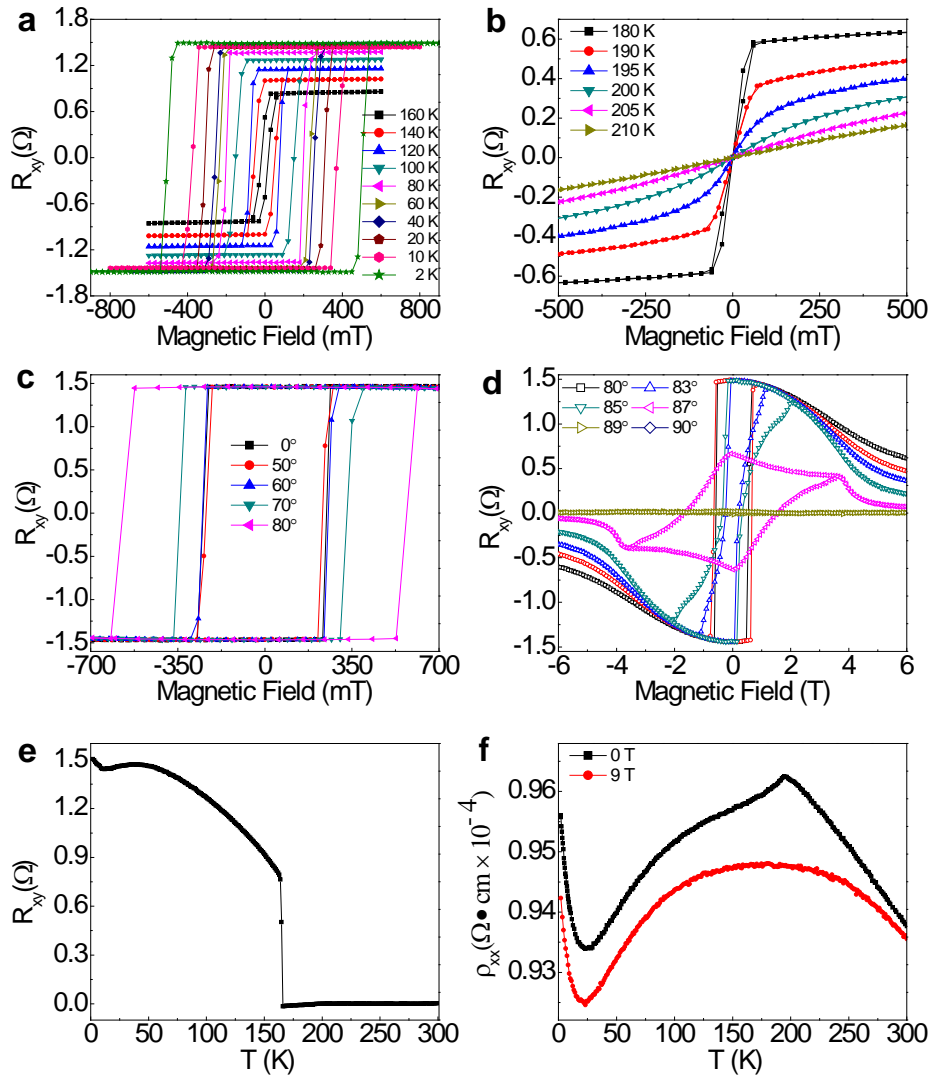

**Supplementary Fig. S6. $R_{xy}$ and $R_{xx}$ measurements for the nanoflake with a thickness of 49 nm.** (a) $R_{xy}$ (B) loops from 2 K to 160 K. (b) $R_{xy}$ (B) loops from 180 K to 210 K. The curve becomes linear at ~210 K. (c-d) Angular dependent Hall effect at 50 K. Loops taken at angles 0° - 50°are not shown, because they nearly overlap. Actually, the coercivity slightly decreases with increasing angles from 10º to 40º, which can still be well fitted using the modified Stoner-Wohlfarth model, as described in part 4 of the supplementary materials. (e) Temperature dependence of remanence. An obvious phase transition occurs at ~ 155 K. When T < ~15 K, the $R_{xy}$ shows an abrupt rise. A phase transition may occur in this temperature regime. (f) The $\rho_{xx}$(T) curve under zero magnetic field displays a peak at ~194 K, attributed to the spin-flip scattering near $T_C$. It can be supressed by a 9 T magnetic field as shown in the figure.

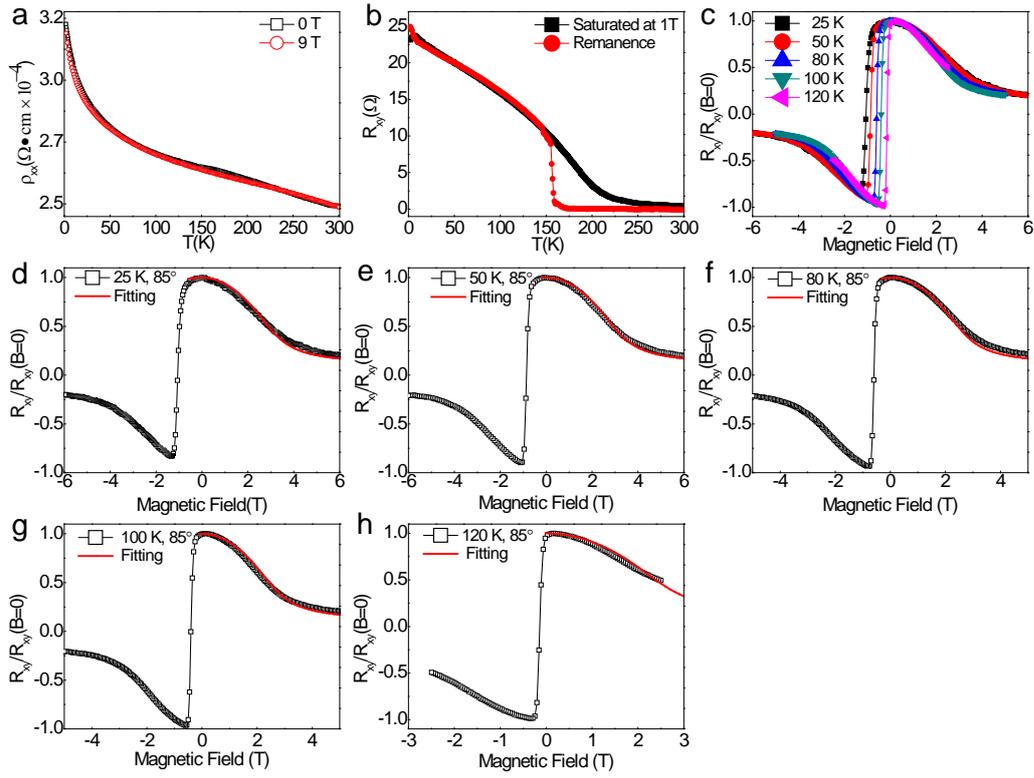

**Supplementary Fig. S7. Temperature dependent $R_{xx}$ and $R_{xy}$ loops from the nanoflake with a thickness of 10.4 nm.** **(a)** Temperature dependence of the $\rho_{xx}$ at 0 T and 9 T. Spin-flip scattering is evident near $T_C$. The enhanced resistance with decreasing temperature may originate from the disorder at the sample surface induced by exfoliation or oxidization. **(b)** Temperature dependence of the remanence $R_{xy}(T)$ and the saturated at 1 T $R_{xy}(T)$ curves. **(c-h)** Normalized angular dependent Hall resistance measured with the magnetic field swept from the positive saturated magnetic field to the negative saturated magnetic field at various temperatures. The red solid lines (d-h) are fitting curves based on the Stoner-Wohlfarth model.

# Section 3: Definition of the Curie temperature

The definition of Curie temperature is a vital part of magnetic material measurements. Generally, based on Curie-Weiss law, we can determine the Curie temperature of a sample through a linear fit to the temperature dependence of inverse magnetization above $T_C$. However, this method is only accurate when the critical exponent is 1 for the temperature-dependent susceptibility, which is unknown yet for FGT nanoflakes. Here we define the temperature at which the remanence $R_{xy}$ goes to zero as $T_C$ [*Nature* **546**, 265–269 (2017)], which can give us more accurate $T_C$ values.

In our experiments, the sample was firstly cooled down to 2 K under a magnetic field of 1 T(-1T). Then the magnetic field was slowly (5 Oe/s) decreased to 0 Oe at 2 K. Finally we scanned temperature from 2 K to 300 K at 3 K/min and get the $R_{xy}$ vs T curve with 1 T (-1 T) remanence. By telling the junction of remanence vs T curve of 1 T and -1 T(Fig. S8b), where $R_{xy}(T)$ goes to zero, we can get the value of $T_C$.

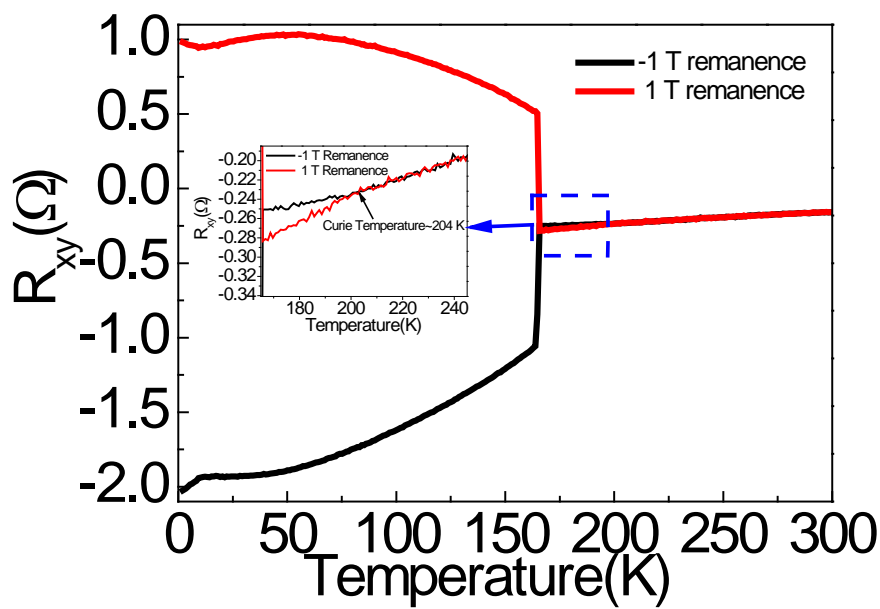

**Supplementary Fig. S8.** Definition of the Curie temperature of one FGT sample. The remanence curves of 1 T and -1 T merge to one curve at ~204 K, which we define as the Curie temperature. At this temperature, the remanence ($R_{xy}(T)$) becomes zero.

# Section 4: Anomalous Hall effect and the Stoner-Wohlfarth model

*Anomalous Hall Effect: Basic formula and data process*

In a ferromagnetic material, the relationship between the Hall resistance and applied magnetic field shows in Eq. S1.

$$R_{xy} = R_0 B_Z + R_S M_Z \tag{S1}$$

Here, $R_{xy}$ is the Hall resistance, which is composed of a normal Hall resistance (the first term in Eq. S1) and an anomalous Hall resistance (the second term in Eq. S1). $B_Z$ and $M_Z$ are the applied magnetic field and the sample magnetic moment perpendicular to the sample surface, respectively. The anomalous Hall effect is proportional to $M_Z$. As FGT is a metallic ferromagnetic material, the normal Hall resistance is very small compared with the anomalous Hall resistance in the magnetic field range of interest. Hence, the shape of the $R_{xy}$ vs B loop is actually the same as that of the $M_Z$ vs B loop when the applied magnetic field is perpendicular to the sample surface. The coercivity and $M_R/M_S$ ratio can be obtained from the $R_{xy}$ (B) curve.

Because of the non-symmetry in our nanoflake devices, the measured Hall resistance was mixed with the longitudinal magnetoresistance. We processed the data by using ($R_{xyA}$(+B) - $R_{xyB}$(-B)) / 2 to eliminate the contribution from the longitudinal magnetoresistance, where $R_{xyA}$ is the half loop sweeping from the positive field to the negative field, $R_{xyB}$ is the half loop sweeping from the negative field to the positive field, and B is the applied magnetic field. We also measured the $R_{xy}$ (T) at the remanence point for most of the samples. To measure the $R_{xy}$ (T) at the remanence point, the magnetic moment of samples was first saturated by a 1 T magnetic field and then the magnetic field was decreased to zero (the remanence point). Finally, the temperature dependence of the $R_{xy}$ at remanence was measured when the temperature was increased from 2 K to 300 K. In order to eliminate the non-symmetry effect of the device, we measured $R_{xy}$ (remanence) with both 1 T and -1 T saturation. The real $R_{xy}$ (T) at remanence without $R_{xx}$ mixing was calculated using ($R_{xyA}$(T) - $R_{xyB}$(T)) / 2. Here $R_{xyA}$ and $R_{xyB}$ are the remanence with 1 T and -1 T saturation, respectively.

*Demagnetization effect*

The demagnetization effect is significant for FGT nanoflakes with perpendicular magnetic anisotropy. The following sentence is from Physical Review B vol 58, 3223, which shows the demagnetization factors for thin film with perpendicular anisotropy. "We approximate the thin film by a homogeneously magnetized ellipsoid of revolution of volume $V$ whose radius $R_x = R_y = R$ is much larger than the 'film thickness' $2R_z$. The magnetostatic self-interaction energy is then given by $D\mu_0 M^2 V/2$, where $D \approx 1$ and $D \approx 0$ are the demagnetizing factors for in-plane and perpendicular magnetization orientations, respectively."

Based on the results above, we can obtain the effective field $B_{eff}$ (magenitude and angle) from the applied magnetic field.

$$B_{eff} = \sqrt{(B\cos\theta)^2 + (B\sin\theta - M|\sin(\phi-\theta)|)^2}$$
$$\theta_{eff} = arctg\left(\frac{B\sin\theta - M|\sin(\phi-\theta)|}{B\cos\theta}\right)$$
(S2)

The schematic diagram of the angular relationship of the applied field, magnetization, and perpendicular anisotropy is shown Fig S9a.

*Theoretical fitting by modified Stoner–Wohlfarth model*

Three different energies, the magnetic anisotropic energy, the Zeeman energy due to the interaction between the applied magnetic field and magnetic moments, and the thermal agitation energy determine the magnetic behavior of the FGT nanoflakes.

*1. Fitting for the $R_{xy}$ hysteresis loops at different temperatures*

From the $M_R/M_S$ ratio (~ 1), we know that all the spins in FGT nanoflakes align perpendicular to the sample surface at the remanence. Except the regime near the coercive field, an FGT nanoflake behaves like a single domain particle. Therefore, we can use the Stoner-Wohlfarth model to describe the magnetic behavior of a FGT nanoflake in the magnetic field regime away from the coercivity. As shown in Fig. S9a, the angles between the applied magnetic field and the direction of the perpendicular anisotropy and the magnetic moment are θ and ϕ, respectively. The direction of the magnetic field means the direction of the POSITIVE magnetic field.

Based on the Stoner-Wohlfarth model, the energy of a FGT nanoflake at temperature T can be written as

$$E(T) = K_A(T)V_S \sin^2(\phi - \theta_{eff}) - M_S(T)B_{eff}V_S \cos\phi \quad \text{(S3)},$$

where $K_A$ is the magnetic anisotropic energy, $V_S$ is the volume of the sample, $M_S(T)$ is the magnetic moment of a unit volume FGT at temperature T, and B is the applied magnetic field. Considering the demagnetization effect as aforementioned, B and θ here should also be modified to $B_{eff}$ and $\theta_{eff}$. With an applied magnetic field B at known θ, we can easily calculate the ϕ value using the equation

$$\frac{\partial E(T)}{\partial \phi} = 2K_A(T)V_S \sin(\phi - \theta_{eff})\cos(\phi - \theta_{eff}) + M_S(T)B_{eff}V_S \sin\phi = 0 \quad \text{(S4)},$$

In anomalous Hall measurements, the $R_{xy}$ is proportional to the value of $M_Z$ (the magnetization perpendicular to the sample surface). Therefore

$$R_{xy}(T) \propto M_S(T)\cos(\phi - \theta_{eff}) \quad \text{(S5)}.$$

By fitting the measured $R_{xy}(T)$ loops at temperature T based on Eq. S4, and Eq. S5, the magnetic anisotropic energy $K_A(T)$ can be obtained. The most accurate $K_A(T)$ value can be fitted based on the loops with θ = 85° as shown in Fig. 4c and Fig. S7 (d-h). After the $K_A(T)$ value is obtained, the angular dependent coercivity can then be fitted based on a modified Stoner-Wohlfarth model at temperature T.

*2. Fitting for the angular dependence of coercivity based on a modified Stoner-Wohlfarth model*

FGT nanoflakes show multi-domain behaviour near the coercivity. In this field regime, the traditional Stoner-Wohlfarth model cannot be utilized. In this case, we modify the Stoner-Wohlfarth model and successfully explain the multi-domain behaviour near the coercivity of FGT nanoflakes.

When the magnetic field is swept to a negative field, the B field in Eq. S3 and S4 is negative. To fit the angular dependence of coercivity, we make *two assumptions*,

1. Eq. S4 can have two kinds of solutions, the stable state (low energy) and the unstable state (high energy state), as shown in Fig. S9b. With an increasing magnetic field B (more negative B value), the energy difference $\Delta E$ between the stable state 1 and the unstable state 2 decreases. At a certain B field, the thermal agitation energy is large enough to overcome the $\Delta E$ in a standard experimental time and the magnetic moment will flip to a stable state in the opposite direction. As the FGT nanoflake shows a nearly square-shaped $R_{xy}$ loop (magnetic loop), we can assume that this B field is the coercive field.

2. When the first domain flips to the opposite direction under an applied magnetic field, other un-flipped magnetic moments will generate an effective field on the magnetic moment in the first flipped domain.

An important issue for the assumption 1 is to determine the ratio $\Delta E/k_B T$ to realize the experimentally observable flipping of domains. Neel and Brown proposed that the relaxation time $\tau$ for the system to reach thermodynamic equilibrium from the saturated state can be

written as

$$\tau^{-1} = f_0 Exp(-\Delta E / k_B T) \tag{S6},$$

where $f_0$ is a slowly variable frequency factor of the order $10^{-9}$ sec$^{-1}$. Assuming $\tau = 100$ Sec, Eq. S6 yields the condition

$$\Delta E = 25 k_B T \tag{S7},$$

Based on Eq. S3, Eq. S4, and Eq. S7, we can write the equations to fit the angular dependence of coercivity at the temperature T.

$$2K_A(T)\sin(\phi_2 - \theta_{eff})\cos(\phi_2 - \theta_{eff}) + M_S(T)B_{eff}\sin\phi_2 = 0 \tag{S8a},$$

$$2K_A(T)\sin(\phi_1 - \theta_{eff})\cos(\phi_1 - \theta_{eff}) + M_S(T)B_{eff}\sin\phi_1 = 0 \tag{S8b},$$

$$[K_A(T)V(T)\sin^2(\phi_2 - \theta_{eff}) - a(T)V(T)M_S(T)B_{eff}\cos\phi_2] - \\ [K_A(T)V(T)\sin^2(\phi_1 - \theta_{eff}) - a(T)V(T)M_S(T)B_{eff}\cos\phi_1] = 25 k_B T \tag{S8c},$$

where V(T) is the volume of the first flipped domain at T, $a$ (T) is the parameter describing the effective field due to the coupling between the first flipped domain and the un-flipped magnetic moments at temperature T, which affects the Zeeman energy. The value of $a$ (T) should be between 0 and 1. $\phi_1$ and $\phi_2$ are the angles between the applied magnetic field and the magnetic moment for the stable state 1 and unstable state 2, respectively, which are NOT fitting parameters. They were calculated using the Eq. S8a and S8b. The domain dynamics in the system is actually very complex. We should not think that the first flipping domain flips to $\phi_2$, while the magnetic moments of other parts of the FGT flake still points to the $\phi_1$ direction. As the magnetic loop of FGT is nearly square-shaped, all the magnetic moments flip and overcome the barrier ($\phi_2$) in a very narrow range of magnetic field. In this situation, using the calculated $\phi_1$ and $\phi_2$ based on Eq. S4 is a pretty good approximation.

V(T) and $a$ (T) are the only two fitting parameters of the three equations in Eq. S8, by which the experimental angular dependence of coercivity at various temperatures can be well fitted , as shown in Fig. 4d. The fitted V(T) and $a$ (T) are also very reasonable. All the $a$ (T) are between 0 and 1, which means the effective field decreases the effect of the Zeeman energy ( $a < 1$ ), while it cannot totally eliminate the effect of Zeeman energy ( $a > 0$ ). The temperature dependence of $a$ (T) shown in Fig. S9d can be explained as below. With increasing temperature, the coercive field decreases and therefore the $\phi - \theta$ decreases. The ratio between the energy due to the coupling of the flipped domain and the un-flipped magnetic moment to the Zeeman energy increases. Hence, $a$ (T) decreases with an increasing temperature.

It should be emphasized that the fitted V(T) and $a$ (T) values are the only possible values to fit the experimental data. Small deviations from the fitting results generate a large difference between the fitting curve and the experimental data(Fig. S9c), which clearly demonstrates that our model provides a reasonable description of the magnetic phenomena in FGT nanoflakes.

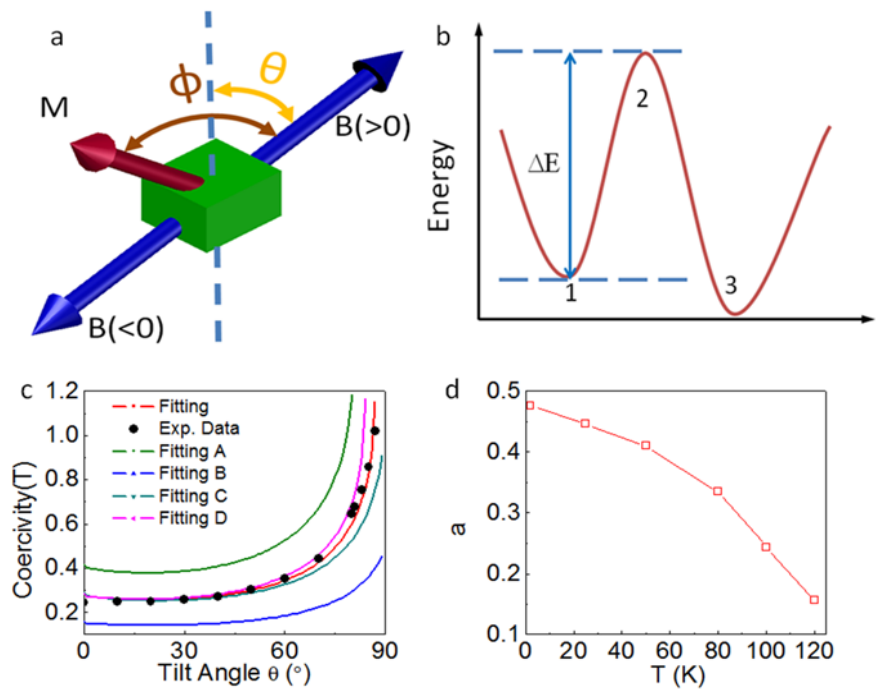

**Supplementary Fig. S9. a**, Schematic of the Stoner–Wohlfarth model. **b**, Schematic diagram of a magnetic system change from stable to unstable state. **c**, Experimental and fitting curves with different values of V and a ($K_A=4.53 \times 10^6$ erg/cc). Fitting: V = 481 nm$^3$, a = 0.407. Fitting A: V = 550 nm$^3$, a = 0.407. Fitting B: V = 430 nm$^3$, a = 0.407. Fitting C: V = 481 nm$^3$, a = 0.46. Fitting D: V = 481 nm$^3$, a = 0.36. **d,** The fitted parameter describing the effective field generated by the interaction between the first flipped domain and the un-flipped magnetic moments.

# Section 5: Mean Field fitting and Spin Wave fitting

To fit the temperature dependent remanence in Fig. 3c, we tried the mean field theory (the Brillouin function) and the spin wave theory.

## The Curie-Weiss mean field theory

Due to the interaction between magnetic moments, an internal field can be written as

$$B_i = n_w M_{sp} \tag{S9}$$

where $n_w$ is a parameter describing the strength of the internal field and $M_{sp}$ is the spontaneous magnetization.

$$M_{sp} = M_0 \Im_J(x) \tag{S10}$$

$$x = \frac{M_0 B_i}{k_B T \cdot N} \tag{S11}$$

Where $M_0$ is the magnetic moment at zero K, $\Im$ is Brillouin function, $k_B$ is Boltzman constant, N the total number of unit magnetic moments. From the above equations, we can obtain

$$\frac{M_{sp}}{M_0} = \Im_J(x) \tag{S12}$$

$$\frac{M_{sp}}{M_0} = \frac{x \cdot N \cdot k_B T)}{n_w M_0^2} \tag{S13}$$

Now we calculate the value of $T_C$

$$n_w M_{sp} \cdot M_0 \approx k_B T_C \tag{S14}$$

Therefore, we obtain

$$n_w M_0 \cdot \Im(x) \cdot M_0 = k_B T_C \tag{S15}$$

We obtain $n_w M_0^2 \frac{(J+1)}{3J} \approx k_B T_C$ (S16)

Combine Eqs S11, S12 and S16, we can calculate the $M_{sp}$ vs ($T/T_C$) curves for different J values.

## Spin wave model

The temperature dependence of magnetic moments of a three dimensional spin wave is

$M \propto a + b(\frac{k_B T}{2J_s})^{3/2}$, which has been discussed in many text books.

# Section 6: The effect of the surface amorphous oxide layer

Although we tried to minimize the exposure of samples to ambient conditions, but a significant oxide layer could still forms quickly on the top of the nanoflakes, which has been confirmed by cross-sectional electron microscopy images shown in Fig. S2a.

We choose the same time of ambient exposure as that throughout our device fabrication and transport measurements (~7 mins). The image shows that there is indeed an amorphous oxide layer of ~1.2 nm thick on the sample surface.

To check whether FGT flakes without oxide layer still show hard magnetic phase with a near square-shaped loop, we fabricated ultraclean devices using our new vdW fabrication system in a glove box with both $O_2$ and $H_2O < 0.1$ ppm. The image of one of the devices is shown in Fig. S10c. To fabricate this device, we utilized the method in [Nature Physics 13, 677–682 (2017)]. Firstly, 5 nm thick Pt contacts were fabricated on Si/SiOx substrate in ambient condition. In our glove box, an exfoliated FGT flake was then dry transferred onto the contacts to form very good ohm contacts. Thereafter, a large hBN flake was dry transferred to cover the FGT flake to prevent oxidizations in measurements. Finally, the sample was covered by PMMA to prevent any possible oxidization. The $R_{xy}$ vs H loop is shown in Fig. S10d. It is very clear that ultra-clean FGT flakes still show hard magnetic property with square shaped loop. The $R_{xy}$ vs T curve also shows the same properties as the device described in the main text. Thus, the thin oxide layer on the sample surface does not affect the main conclusions (hard magnetic properties with a near square-shaped loop) of this paper. Moreover, we can see that FGT is a promising material whose magnetism can survive in ambient environment for a certain time.

From our experiments, we conclude that the effect of oxide layer includes:

1.      The switch of magnetic moment in the square shape loop of FGT with oxide layer is not as sharp as that in ultra clean FGT flakes, which is due to the pinning effect of the oxide layer.

2.      The coercivity of FGT slightly increases after the oxidization, which is also due to the domain wall pinning effect.

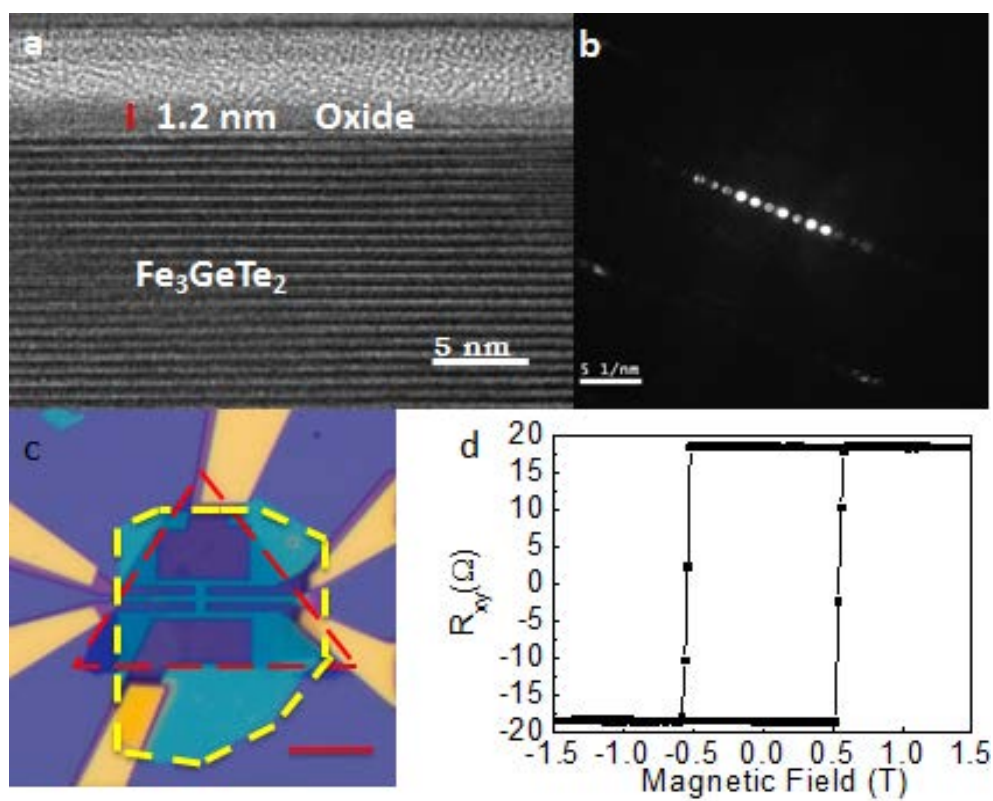

**Supplementary Fig. S10** (a) Cross-sectional TEM image of an FGT nanoflake on substrate. The top layer is 5 nm Pt layer. The oxide layer is about 1.2 nm. The thickness of monolayer is about 0.8 nm. The scale bar is 5 nm. (b) Diffraction pattern of the FGT nanoflake. (c) A 5.8 nm FGT device covered by h-BN, the bottom contact is 5 nm Pt. The red dashed line is FGT region, yellow dashed line is h-BN region. The scale bar is 10 μm. (d) Anomalous Hall effect at 2 K for this device. Magnetic field is perpendicular to the sample surface.

# Section 7: The confirmation of ohmic contact

We fabricated an FGT device using the same recipe as all the other samples and confirm that our procedure of device fabrication produces good ohmic contacts.

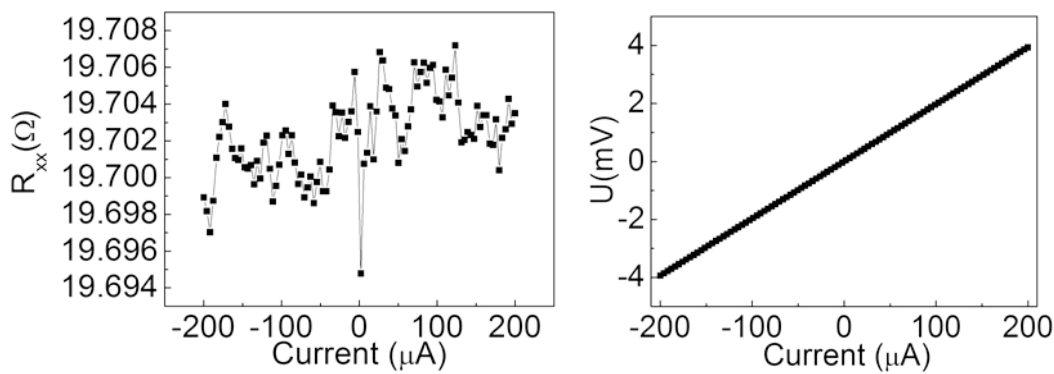

**Fig. S11** $R_{xx}$ vs Current curve at 2 K for an FGT sample. (b) Corresponding I-V curve derived from (a).

# Supplementary References